\documentclass[twocolumn,groupedaddress,amsmath,epsfig,amssymb]{revtex4}
\usepackage{makecell}
\usepackage{bbm}
\usepackage{mathrsfs}
\usepackage{amsfonts}
\usepackage{color}
\usepackage{graphicx}
\newcommand{\dbar}{d\hspace*{-0.08em}\bar{}\hspace*{0.1em}}
\begin{document}
\newcommand{\Rv}{{\vec {R}}} 
\newcommand{\rv}{{\vec r}}
\newcommand{\tv}{{\vec t}}
\newcommand{\av}{\boldsymbol a}
\newcommand{\fv}{{\boldsymbol f}}
\newcommand{\mv}{{\boldsymbol m}}
\newcommand{\nv}{{\boldsymbol n}}
\newcommand{\hv}{{\boldsymbol h}}
\newcommand{\xv}{{\boldsymbol x}}
\newcommand{\xp}{\vec{x}_{\perp}}
\newcommand{\pp}{\vec{p}_{\perp}}
\newcommand{\zv}{{\boldsymbol z}}
\newcommand{\uv}{{\vec u}}
\newcommand{\Av}{{\boldsymbol A}}
\newcommand{\Xv}{{\boldsymbol X}}
\newcommand{\Yv}{{\boldsymbol Y}}
\newcommand{\Pv}{{\boldsymbol P}}
\newcommand{\Qv}{{\boldsymbol Q}}
\newcommand{\Hv}{{\boldsymbol H}}
\newcommand{\ur}{\vec{{\EuFrak u}}}
\newcommand{\cv}{{\vec c}}
\newcommand{\qv}{{\boldsymbol q}}
\newcommand{\pv}{{\boldsymbol p}}
\newcommand{\yv}{{\boldsymbol y}}
\newcommand{\vv}{{\boldsymbol v}}
\newcommand{\kv}{{\vec k}}
\newcommand{\phiv}{{\boldsymbol \phi}}
\newcommand{\etav}{{\boldsymbol \eta}}
\newcommand{\Tr}{{\rm Tr}}
\newcommand{\px}{{\partial_x}}
\newcommand{\py}{{\partial_y}}
\newcommand{\ppi}{{\partial_i}}
\newcommand{\ppj}{{\partial_j}}
\newcommand{\ch}{{\hat{c}}}
\newcommand{\eh}{{\hat{e}}}
\newcommand{\xh}{{\hat{x}}}
\newcommand{\yh}{{\hat{y}}}
\newcommand{\zh}{{\hat{z}}}
\newcommand{\vh}{{\hat{v}}}
\newcommand{\qh}{{\hat{q}}}
\newcommand{\kh}{{\hat{k}}}
\newcommand{\llm}{{\boldsymbol \lambda}}
\newcommand{\Am}{{\boldsymbol{A}}}
\newcommand{\Qm}{{\boldsymbol{Q}}}
\newcommand{\Rm}{{\boldsymbol R}}
\newcommand{\Lm}{{\boldsymbol L}}
\newcommand{\Km}{{\boldsymbol K}}
\newcommand{\Jm}{{\boldsymbol J}}
\newcommand{\Tm}{{\boldsymbol T}}
\newcommand{\Bm}{{\boldsymbol B}}
\newcommand{\Dm}{{\boldsymbol D}}
\newcommand{\Cm}{{\boldsymbol C}}
\newcommand{\Em}{{\boldsymbol E}}
\newcommand{\Mm}{{\mathcal M}}
\newcommand{\Wm}{{\boldsymbol W}}
\newcommand{\Fm}{{\boldsymbol F}}
\newcommand{\Gm}{{\boldsymbol G}}
\newcommand{\Imm}{{\boldsymbol I}}
\newcommand{\sm}{{\boldsymbol s}}
\newcommand{\gammam}{{\boldsymbol \Gamma}}
\newcommand{\chim}{{\boldsymbol \chi}}
\newcommand{\be}{\begin{equation}}
\newcommand{\ee}{\end{equation}}
\newcommand{\ba}{\begin{eqnarray}}
\newcommand{\ea}{\end{eqnarray}}
\newcommand{\RNum}[1]{\uppercase\expandafter{\romannumeral #1\relax}}
\newcommand{\ddelta}{\boldsymbol{\delta}}
\newcommand{\pdf}{\mathcal{P}}
\newcommand{\U}{\mathcal{U}}
\newcommand{\LFP}{\mathcal{L}_{\rm FP}}

\title{Hamiltonian Heat Baths, Coarse-Graining and Irreversibility:\\ A Microscopic Dynamical Entropy from Classical Mechanics}
\author{Mingnan Ding$^{1,2}$}
\email{dmnphy@sjtu.edu.cn}
\author{Michael E. Cates$^{1}$}
\affiliation{$^1$DAMTP, Centre for Mathematical Sciences, University of Cambridge,
Wilberforce Road, Cambridge CB3 0WA, United Kingdom\\
$^2$Wilczek Quantum Center, School of Physics and Astronomy, Shanghai Jiao Tong University, Shanghai, 200240 China\\
}

\date{\today}

\begin{abstract} 
The Hamiltonian evolution of an isolated classical system is reversible, yet the second law of thermodynamics states that its entropy can only increase. This has confounded attempts to identify a `Microscopic Dynamical Entropy' (MDE), by which we mean an entropy computable from the system's evolving phase-space density $\rho(t)$, that equates {\em quantitatively} to its thermodynamic entropy $S^{\rm th}(t)$, both within and beyond equilibrium. Specifically, under Hamiltonian dynamics the Gibbs entropy of $\rho$ is conserved in time; those of coarse-grained approximants to $\rho$ show a second law but remain quantitatively unrelated to heat flow. Moreover coarse-graining generally destroys the Hamiltonian evolution, giving paradoxical predictions when $\rho(t)$ exactly rewinds, as it does after velocity-reversal. 
Here we derive the MDE  for an isolated system XY in which subsystem Y acts as a heat bath for subsystem X. 
We allow $\rho_{XY}(t)$ to evolve without coarse-graining, but compute its entropy by disregarding the detailed structure of $\rho_{Y|X}$. The Gibbs entropy of the resulting phase-space density $\tilde\rho_{XY}(t)$ comprises the MDE for the purposes of both classical and stochastic thermodynamics.  The MDE obeys the second law whenever $\rho_X$ evolves independently of the details of Y, yet correctly rewinds after velocity-reversal of the full XY system. 
\end{abstract}
\maketitle

The second law of thermodynamics tells us that the thermodynamic entropy $S^{\rm th}(t)$ of an isolated system can increase with time $t$, but not decrease. In particular, this holds for  $S^{\rm th}_{XY}(t)$ in a system X coupled to a heat bath Y, such that the XY system is isolated. Increases in $S^{\rm th}$ encode the irreversibility of thermal processes, and contrast with the reversibility of the underlying Hamiltonian dynamics. How to view the emergence of irreversibility has been discussed for over a century~\cite{Thomson1874,Brush1966,Wu1975,Lebowitz2005,teVrugt2021}. The development of stochastic thermodynamics (reviewed in~\cite{Seifert2012}) has brought renewed importance to this debate, and recent progress in Hamiltonian formulations of stochastic thermodynamics~\cite{Evans2001,Evans2002,Ding2022,Ding2022-1} have brought into focus the lack of an agreed statistical definition of $S^{\rm th}_{XY}(t)$. 

The second law originated from the study of heat engines~\cite{Thomson1874}. Its key insight is that the entropy change of the heat bath Y at temperature $T$ is $-\dbar Q/T$, with $-\dbar Q$ the heat flow from X to Y~\cite{Carnot1872,Clausius1854,Clausius1865}. This does not require the system X to be large, or near equilibrium, or to have a well-defined temperature. ($T$ will be the bath temperature below; we set $k_B=1$.)  To help understand irreversibility's microscopic origins, Boltzmann defined the microcanonical entropy for an isolated system of energy $E$ as $S^\Omega = \log \Omega(E)$ with $\Omega$ its phase-space volume~\cite{Ehrenfest1959,Carlo2006,Boltzmann2012,Lebowitz2005}. However, his subsequent H-theorem does not explain irreversibility since its assumptions already break time-reversal symmetry~\cite{Burbury1894,Bryan1894}. More generally, there is no agreed definition of $\Omega$ out of equilibrium; attempts based on an  `accessible' volume contradict experiments~\cite{Cates2015}.

The Gibbs entropy $S^G = - \int \rho \log \rho$,  describing an ensemble of systems with phase-space density $\rho$, is widely viewed as a microscopic counterpart of $S^{\rm th}$~\cite{Gibbs1902,Jaynes1965}. (The von Neumann entropy in quantum theory~\cite{Neumann2018} and the Shannon entropy in information theory~\cite{Shannon1948,Thomas2006,Gell-Mann1996} are both closely related to $S^G$.)  But although it matches $S^{\rm th}$ in equilibrium~\cite{Jaynes1965},  $S^G$ is conserved for isolated Hamiltonian systems. Thus $S^{\rm th}_{XY}(t)\neq S^G_{XY}(t)$. 

A standard approach, dating back to Gibbs himself, is coarse-graining: phase space is partitioned into cells, within which $\rho$ is repeatedly replaced by its cell-wise average $\bar\rho$~\cite{Gibbs1902,Ehrenfest1959,Maes2003,teVrugt2021}. Because $S^G[\bar\rho]\ge S^G[\rho]$ for any function $\rho$, the coarse-grained Gibbs entropy $S^{\rm cg}(t) = S^G[\bar \rho]$ is strictly non-decreasing in time. There are two problems with this approach, both well known. First, without a clear prescription on how cells are chosen, it offers no {\em quantitative} theory of heat flow as classical thermodynamics demands. It also fails to connect with stochastic thermodynamics in which, at trajectory rather than ensemble level, the second law is replaced by quantified small probabilities of $S^{\rm th}$ decreasing in time~\cite{Seifert2012}. 

A second  problem is that $S^{\rm cg}$ remains strictly non-decreasing in laboratory experiments, and/or thought-experiments, that suddenly reverse of all the velocities (strictly, generalized momenta) in the XY system so as to {\em exactly rewind} the prior trajectory. This is paradoxical since for an evolved-and-reversed system, $S^{\rm cg}_{XY}$ takes different values in initial and final states that are {\em exactly} the same. This `echo paradox', named after the spin-echo experiments that inspired it~\cite{Ridderbos1998}, is discussed later. 

This paper constructs a Microscopic Dynamical Entropy (MDE), $S^{\rm md}_{XY}$, from the phase-space density $\rho_{XY}$  of the XY system by a different coarse-graining approach that preserves reversibility. Our MDE quantitatively matches the entropy $S^{\rm th}_{XY}$ of both classical and stochastic thermodynamics. It also resolves the echo paradox.

Our MDE is essentially the sum of the Gibbs entropy $S^G_X = -\int\rho_X({\xv})\ln\rho_X({\xv})$ of subsystem X, and the mean (over $\rho_X$) of the microcanonical entropy $S_Y^\Omega$ of its heat bath Y. Here $S_Y^\Omega$ depends on $\rho_X$ only through the X-subsystem energy, $E_X $. This MDE is constructed by `nondestructive' coarse-graining: details of the conditional density $\rho_{Y|X}$, beyond those implied by energy conservation, are disregarded when calculating the Gibbs entropy -- though {\em not} when evolving $\rho_{XY}(t)$. We show that $S^{\rm md}_{XY}$ depends on $\rho_X$ only. Our construction directly mirrors the thermodynamic idea that a system and its heat bath communicate only by exchange of heat energy.

We will show that $S^{\rm md}_{XY}(t) = S^{\rm th}_{XY}(t)$ whenever Y is large enough to be a heat bath. We will also establish a second law for $S^{\rm md}_{XY}$, but only when the influence of Y on the evolution of $\rho_X$ is `heat-bath-like'. That is, the X evolution must depend only on a few parameters of Y, not its details. Thus the second law does not apply in echo protocols, where velocity details in Y are paramount.

Our coarse-graining strategy can be motivated in part by time-scale separation, with a fast-relaxing heat bath Y and slower dynamics for X. Notably though, such a separation does not remove information, and $S^G_{XY}$ is still constant. 
This is compatible with an increasing $S^{\rm md}_{XY}$ because the latter ignores the structure of $\rho_{Y|X}$, whose ever-increasing complexity compensates the rising $S^{\rm md}_{XY}(t)$, and sustains the constancy of $S^G_{XY}$. 

\begin{figure*}
\begin{centering}
\includegraphics[width=2\columnwidth]{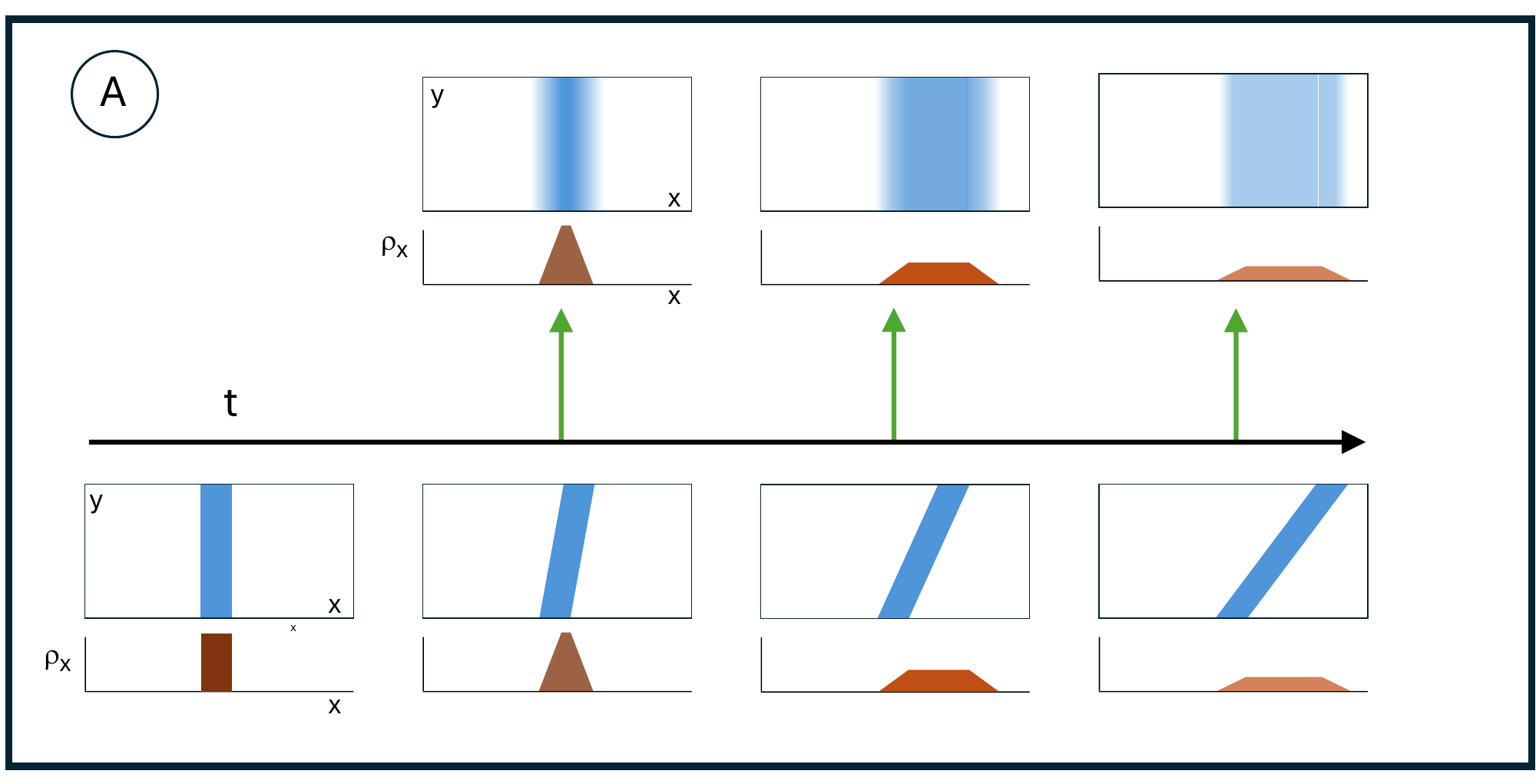}\\
\includegraphics[width=2\columnwidth]{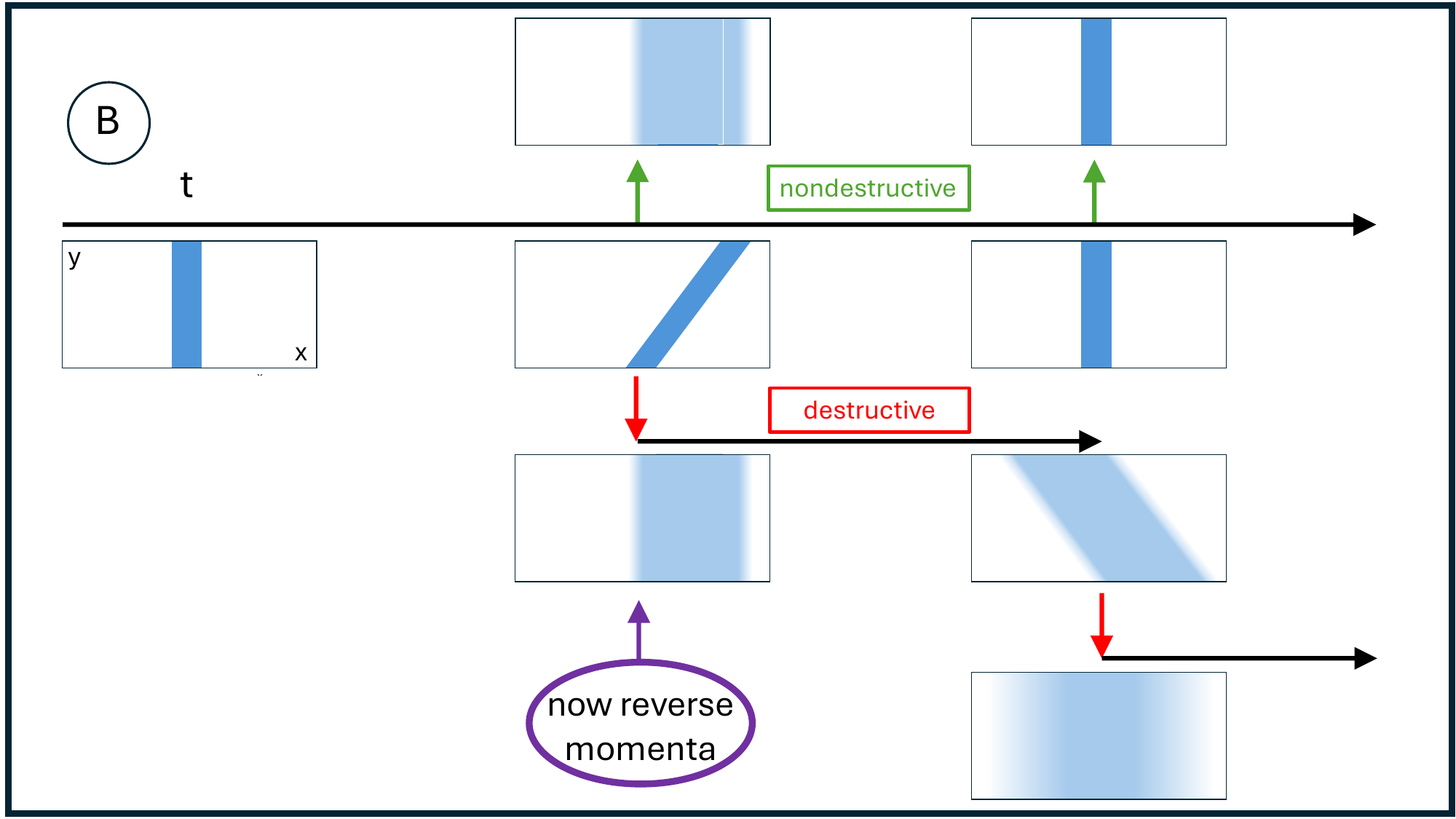}
\par\end{centering}
\caption{Schematic evolution of full and coarse-grained densities in phase space. {\bf (A)} Phase space of the XY system is represented schematically as a rectangle with only one coordinate for each subsystem. The height of the rectangle (range of the $y$ coordinate) is $\Omega_Y$. This is constant for simplicity here, corresponding to the case where all states $x$ of subsystem X have equal $E_X$. The lower row of panels in (A) show the evolution of the full density $\rho_{XY}$ in blue, and below it the marginal $\rho_X$ in brown. Initially -- leftmost panel  -- subsystems Y and X are uncorrelated (hence the density is uniform in $y$ whenever nonzero in $x$). The time evolution (horizontal timeline) is represented as a shearing within the XY phase space, which conserves local densities, as Liouville's theorem requires. The lower row shows the exact evolution, with ever-increasing dependence of $y$ on $x$.  The vertical green arrows show our chosen coarse-graining procedure: at each time-slice, the density $\rho_{Y|X}$  is homogenized over the $y$ coordinate, leaving $\rho_X$ unaffected. {\bf (B)} Comparison of destructive and nondestructive coarse graining in an `echo' (momentum reversal) protocol. Leftmost panel: the same initial state as in (A). This evolves to the same final state as in (A), now the central one on the horizontal timeline. Momenta are then reversed and the initial state later reappears (echo effect) in the rightmost panel. Vertical green arrows show the same coarse-graining as in (A). This is nondestructive since it doesn't change the evolution of $\rho_{XY}$; it is {\em only} used in the entropy calculation. The echo effect is preserved. The downward red arrows show a destructive coarse-graining scheme that we do not use, but which (for these purposes) resembles the one considered in Gibbsian phase-cell coarse-graining to find $S^{\rm cg}(t)$. Each red arrow coarse-grains the density but this is then used for subsequent time evolution, creating irreversible dynamics that destroys the echo. A similar effect could also be achieved by replacing the bath Y with a fresh one, uncorrelated with X, just after reversal.}   \label{fig1}
\end{figure*}

These arguments resemble recent treatments of XY systems at quantum level~\cite{Esposito2010,Bera2017,Manzano2018,Li2019,Ptaszynski2019,Landi2021}, but we have not seen one from which our MDE emerges as the classical limit. In particular, we do not use any approximation in which the marginal $\rho_Y$ has energy fluctuations that violate the first law, $E_Y = E-E_X$~\cite{Esposito2010,Li2019}. (This is crucial when X is small so that the bath is almost microcanonical.) Therefore while our {\em schema} for explaining the second law, shown in Fig.~\ref{fig1}, superficially resembles those considered in~\cite{Esposito2010,Li2019}, the results given below are new. These provide a {\em quantitative} microscopic interpretation of the thermodynamic entropy $S^{\rm th}_{XY}(t)$ in isolated XY systems obeying classical mechanics.
\\

\noindent{\bf Coupling via Hamiltonian of mean force:}

We consider a Hamiltonian system XY where Y will later become a heat bath for X. Since thermodynamics is not restricted to weakly coupled systems, we do not assume this here. Instead, we use a formalism for strong coupling established in~\cite{Ding2022}, which also explains its relation to other approaches. 

In general the total Hamiltonian takes the form
\ba
H_{XY}(\xv,\yv) = H^0_X(\xv)+ H^0_Y(\yv)+ H_I(\xv,\yv)\,,
\ea
where $\xv,\yv$ denote the degrees of freedom (both positions and momenta) of X and Y respectively. 
Since Y is later to be viewed as a heat bath, we need to split $H_{XY}$ into two parts, not three: one for the system X and the other for its environment Y. Following~\cite{Ding2022} we write 
\ba
H_Y(\xv,\yv) &=& H_{XY}(\xv,\yv) -H_X(\xv) 
\label{split-H-0}\\
H_X({\xv}) &=& - T \log \frac{ \int  e^{-\beta H_{XY}} {d \yv}  }{ \int e^{-\beta H^0_Y} {d \yv}  }\,.
\ea
$H_X$ is the Hamiltonian of mean force~\cite{Jarzynski2004,Kirkwood1935,Campisi2009}. Here $\beta$ is undetermined but will shortly be identified as the inverse bath temperature. This partitioning has the advantage that almost all thermodynamic relations take the same form
as they do in weak coupling~\cite{Ding2022}. 

As a result of the above split, the partition function $Z_Y=\int { e^{- \beta H_Y(\xv,\yv) } d \yv}$ does not depend on $\xv$~\cite{Ding2022}. 
It decomposes as 
\begin{equation}
Z_Y = \int \Omega_Y(E_Y) e^{-\beta E_Y}  dE_Y\,, \label{eq:laplace}
\end{equation}
with microcanonical phase volume $\Omega_Y$ and entropy $S^\Omega_Y$:
\ba
\Omega_Y(E_Y) =e^{S^\Omega_Y(E_Y)} \equiv \int \delta (H_Y(\xv,\yv) - E_Y)d\yv \,.
\ea
This depends on $\xv$ solely via $E_Y = E-H_X({\xv})$ with $E$ the total energy of the combined XY system.

We now assume that Y acts as a large heat bath (although these arguments can be generalized to consider a small one). In that case we recover the usual thermodynamic relation (again, independent of $\xv$)
\ba
\beta = \frac{ \partial S^\Omega_Y}{ \partial E_Y} \bigg|_{E_Y=E} \,.
\label{beta-def-1}
\ea
\\
\noindent{\bf Microscopic Dynamical Entropy:}

The XY system is described by its phase-space density $\rho_{XY}(\xv,\yv)$, which evolves by Hamiltonian dynamics under $H_{XY}$. The Gibbs entropy of the coupled system 
\ba
S^G_{XY} = - \int \rho_{XY}(\xv,\yv) \log \rho_{XY}(\xv,\yv) d\xv d\yv
\label{Gibbs-entropy}
\ea
is then constant in time via Liouville's theorem~\cite{Landau1976,Sethna2021}. 

We now define the MDE as
\ba
S^{\rm md}_{XY} &\equiv &S^G_X - \beta \langle E_X \rangle + S^\Omega_Y(E)\,,
\label{NTE-0}
\ea
which involves $\rho_X(\xv) =  \int \rho(\xv,\yv) d\yv$, the marginal density in X,  its Gibbs entropy $S^G_X$, and its mean energy $\langle E_X\rangle$ (all of them time-dependent):
\ba
S^G_X &=& - \int \rho_X(\xv) \log \rho_X(\xv)d\xv;\\
\langle E_X\rangle &=& \int \rho_X(\xv) H_X({\xv})d\xv\,.
\ea
In classical thermodynamics, up to the additive constant  $S^\Omega_Y(E)$, $S^{\rm md}_{XY}$ in \eqref{NTE-0} equates to $-A_X/T$ where the availability is $A \equiv E - TS$ for a system at fixed volume in contact with a heat bath at temperature $T$~\cite{Pippard1966}. Indeed, $-A_X/T$ is the total thermodynamic entropy $S^{\rm th}_{XY}$ in our setup, and the second law then states $\dot A_X \le 0$. Near equilibrium, when a definite temperature can be assigned to X itself and not just to the bath Y, $A_X$ becomes the Helmholtz free energy $F_X = E_X-TS_X$~\cite{Pippard1966}. 

Our argument that $S^{\rm md}_{XY} = S^{\rm th}_{XY}$ is further confirmed by considering the MDE production rate which obeys
\ba
dS^{\rm md}_{XY} &=& dS^G_X - \beta \dbar Q_X \\
\Rightarrow\dot S^{\rm md}_{XY}& = &\dot S^G_X - \beta \int_X \left (\partial_t  \rho_{X}  \right) H_Xd\xv  \,.
 \label{s-tot-0}
\ea
This coincides with the  `total entropy production' as defined in stochastic thermodynamics (ST)~\cite{Seifert2012}. 

ST asserts that the entropy production $\Delta\mathcal{S}$ on a path $\gamma$ is the  log-ratio of probability densities, $\ln(\mathbb{P}(\gamma)/\mathbb{P}(\gamma_\mathcal{T}))$, for $\gamma$ and its time reversal $\gamma_\mathcal{T}$~\cite{Seifert2005,Seifert2012,Lebowitz1999,Jarzynski2011,Batalhao2015}. The entropy $\mathcal S$ in question is the sum of the Gibbs entropy of the studied system, X, and $-\dbar Q/T$ for its environment, Y; this is a pathwise version of our $S^{\rm md}_{XY}$. 
Indeed, for Hamiltonian systems without velocity-dependent external forces, Refs.~\cite{Ding2022,Ding2022-1} showed that in infinitesimal processes
\ba
\langle d\mathcal S\rangle = dS^G_X - \beta \dbar Q = dS^{\rm md}_{XY}\,
\ea
with $\langle.\rangle$ the average over trajectories.
\\

\noindent{\bf Statistical interpretation:}

We now elucidate the difference between the Gibbs entropy $S^G_{XY}$ and our MDE by writing 
\ba
\rho_{XY}(\xv,\yv,t) = \rho_X(\xv,t) \rho_{Y|X}(\xv,\yv,t)
\ea
with $\rho_{Y|X}$ the conditional distribution for $\yv$ given $\xv$. If there is time-scale separation between a fast bath Y and a slow X subsystem, then at intermediate time-scales, the marginal $\rho_X$ does not undergo significant changes while $\yv$ fluctuates constantly. The conditional distribution $\rho_{Y|X}(\xv,\yv,t)$ then describes this dynamics. 

The evolution of $\rho_{Y|X}$ is crucial for conservation of the total Gibbs entropy. Indeed, since
\ba
S^G_{XY} & = & S^G_X + S^G_{Y|X}; \\
S^G_{Y|X}& = &- \int \rho_X \rho_{Y|X} \log \rho_{Y|X} d\xv d\yv\,,
\ea
the difference between the Gibbs entropy and the MDE is the replacement of $dS_{Y|X}$ with $-\beta \dbar Q$. It is only by virtue of $S^G_{Y|X}$, and the correlations that it describes between Y and X, that $S^G_{XY}$ remains constant. In contrast, $S^{\rm md}_{XY}$ is determined by $\rho_X$ alone and contains no information about $\rho_{Y|X}$. Therefore when $S^{\rm th}_{XY}$ increases under the second law, this is because it disregards {\em increasingly complex correlations} between the bath Y and the subsystem X, as encoded in  $\rho_{Y|X}(\xv,\yv,t)$. 

We next show that $S^{\rm md}_{XY}$ is the Gibbs entropy $S^G_{XY}[\tilde\rho]$ after the following coarse-graining (Fig.~\ref{fig1}A):
\ba
\tilde \rho_{XY} &=& \rho_X \tilde \rho_{Y|X}(\xv,\yv) ; \nonumber\\
\tilde \rho_{Y|X}(\xv,\yv) & = &
\left \{ \begin{array}{l}
\frac{1}{ \Omega_Y(E_Y)}, \,\, {\rm for} \,\, H_Y = E_Y({\xv})
\\ 
0, { \rm otherwise}
\end{array} \right. \,.
\label{cg-00}
\ea
To establish this, 
write
\ba
&&S^G_{XY}[\tilde \rho_{XY}]  =  \int \rho_X\tilde\rho_{Y|X} \log(\rho_X\tilde\rho_{Y|X})d{\xv}d{\yv} \nonumber\\
& = & S^G_X - \int  \rho_Xd\xv  \int_{H_Y = E_Y} \frac{ 1}{ \Omega_Y(E_Y)} \log  \frac{ 1}{ \Omega_Y(E_Y)} d\yv \nonumber\\
& = & S^G_X + \int  \rho_X \log   \Omega_Y(E_Y)d\xv \nonumber\\
& = & S^G_X +  \int \rho_X  S^\Omega_{Y}(E - E_X(x)) d\xv \nonumber\\
& \simeq & S^G_X +  \int \rho_X  \left ( S^\Omega_Y(E )- \beta E_X(x) \right)d\xv  \nonumber\\
& = & S^G_X - \beta \langle E_X \rangle + S^\Omega_Y(E ) \equiv S^{\rm md}_{XY} [ \rho_{XY}]\,.
\label{cg-entropy}
\ea
In the penultimate line, \eqref{beta-def-1} was used to expand $S_Y^\Omega$ for $E_X\ll E$. The constant $S^\Omega_Y(E)$ was introduced in \eqref{NTE-0}. 

Quantitatively, therefore, our MDE leaves the ${\xv}$ variables intact, but coarse-grains away the details of the bath degrees of freedom $\yv$, replacing the latter with a uniform phase space density at energy $E_Y=E-E_X$. As stated previously, this coarse-graining only applies to the entropy calculation, not to the full dynamics.
\\

\noindent{\bf  Emergence of the Second Law:}

We cannot expect a second law to emerge for $S^{\rm md}_{XY}$ under strict Hamiltonian evolution of $\rho_{XY}$, due to Poincar\'e recurrence which restores reversibility over unphysically long time scales (the Zermelo paradox) ~\cite{Wu1975,Davies1994}. Strictly, the MDE should increase most of the time with negative excursions that, for a sufficiently large systems and/or time-scale separation, are exceedingly rare ~\cite{Davies1994,Linden2009,Strasberg2017,Pavliotis2008}. To get a second law, these must be suppressed by an additional approximation. 

To achieve this we assume that the influence of Y on $\rho_X$ is truly `bath-like': it is conveyed by a sub-extensive set of parameters (such as $T$ and some friction coefficients) rather than by the precise state of Y. More formally, we assume the evolution $\rho_X$ is autonomous in the following sense: if two distributions $\rho_{XY}^{(1)}$ and $\rho_{XY}^{(2)}$ share the marginal
$
\rho^{(1)}_{X}(t_0) = \int   \rho^{(1)}_{XY}(t_0)d\yv
=  \int  \rho^{(2)}_{XY}(t_0)d\yv = \rho^{(2)}_{X}(t_0) 
$
at initial time $t_0$, then for all $t>t_0$:
\ba
\rho^{(1)}_X(t) = \rho^{(2)}_X(t) + O(\epsilon)\simeq \rho^{(2)}_X(t)\,. 
\label{bath-req-2}
\ea
Here $\epsilon$ is a small parameter, in some cases precisely identifiable via a time-scale separation between X and Y~\cite{Ding2023}. The $O(\epsilon)$ correction describes Poincar\'e recurrence and similar effects, which we now suppress.

To prove the second law from \eqref{bath-req-2}, consider an initial density $\rho_{XY}(t_0)$ and its coarse-grained form $\tilde\rho_{XY}(t_0)$ obeying \eqref{cg-00}. By construction these have the same MDE, $S^{\rm md}_{XY} = S^G_{XY}[\tilde\rho_{XY}(t_0)]$. Now evolve $\tilde\rho_{XY}(t_0)$ under $H_{XY}$ to time $t>t_0$. The Gibbs entropy of $\tilde\rho_{XY}$ remains constant as it evolves into $\rho^*_{XY}(t)$ which {\em no longer} obeys \eqref{cg-00}. Now coarse-graining a second time  gives
\ba
S^{\rm md}_{XY }[ \rho^*_{XY}(t)]&\equiv& S^G_{XY }[ \tilde \rho^*_{XY}(t)] \nonumber\\
\geq S^G_{XY }[ \rho^*_{XY}(t)] &=& S^{\rm md}[\rho_{XY}(t_0)].\label{coarse}
\ea
The inequality holds because replacing $\rho^*_{Y|X}$ by a uniform distribution (in $\yv$, see \eqref{cg-00}) cannot decrease its entropy.

We now invoke \eqref{bath-req-2} to say that $\rho^*_{XY}(t)$ has the same $\rho_X(t)$ as $\rho_{XY}(t)$ and hence the same $S^{\rm md}_{XY}$. Here $\rho_{XY}(t)$ is the time-evolved full density. The second law
\ba
S^{\rm md}_{XY} [  \rho_{XY}(t)] \geq S^{\rm md}_{XY} [ \rho_{XY}(t_0)] \label{SecondL}
\ea
then follows. 
Note that we did not assume equilibrium at the initial and/or final time~\cite{Jaynes1965}. Our only assumption is to omit the ``$O(\epsilon)$" part of \eqref{bath-req-2}. Precise identification of $\epsilon$ may, in general, need more work case-by-case.

Note that under the assumption that the dynamics of X is autonomous, it is in principle permissible to perform coarse-graining in Eq.~(\ref{cg-00}) at each time step destructively. In this sense, the second law is a justification for destructive coarse-graining and not {\em vice versa}. (This might be viewed as assuming perfect time-scale separation in which Y relaxes immediately to a typical state, given X). However, destructive coarse-graining is best avoided even here, as it will give incorrect statistics for the $O(\epsilon)$ terms.\\

\noindent{\bf Resolving the echo paradox:}

Consider an isolated box of gas with a partition that is removed at $t=0$ with the gas in equilibrium in one half of the container. Its thermodynamic entropy $S^{\rm th}(t)$ increases as the gas expands. If at $t=t_r$ velocities are reversed, then at $t=2t_r$ the partition can be re-inserted with zero work to recreate (modulo a second velocity reversal) the exact initial state. This, with the partition restored, is {\em again an equilibrium state}. So, $S^{\rm th}$ has now gone down again, because it is a state function. Any microscopic definition of entropy whose second law still holds after velocity-reversal {\em denies} thermodynamics, rather than confirming it as sometimes supposed~\footnote{While the paradox is not usually framed for an XY system it applies equally to these, {\em e.g.}, an isolated system containing colloids (X) in a solvent bath (Y) with the colloids confined by a semipermeable barrier. The Y momenta must then be reversed alongside those of X.}.

The phase-cell approach  fails in echo protocols because its coarse-graining alters the dynamics of $\rho$ (Fig.~\ref{fig1}B). Even if the cell size it taken to encode real resolution limits, this contradicts the precept that in classical mechanics, observations cannot increase phase-space uncertainty. Similar objections apply to other dynamical approximations involving projection operators, dynamic density functionals, Fokker-Planck equations, {\em etc.}~\cite{Zwanzig2001,teVrugt2021}. These also do not describe the dynamics after velocity-reversal, because they also break reversibility. 

The echo paradox (which is a version of Loschmidt's paradox~\cite{Wu1975}) is linked to a wider set of `irreversibility problems', informatively surveyed in~\cite{teVrugt2021}. Concepts such as `fine-grained equilibrium' have been introduced to escape it, alongside suggestions that $S^{\rm th}$ only increases through  isolation-violating `interventions'~\cite{Ridderbos1998}. 

The paradox applies to all isolated systems, including those of XY type studied here. For these it is resolved by our MDE, which is time-local functional of the true phase-space density (Fig.~\ref{fig1}B). Accordingly, for equilibrium densities, however reached, it is a state function. Moreover its second law \eqref{SecondL} requires autonomy of subsystem X, allowing violations when the influence of Y on X is not `bath-like'. Y clearly ceases to be bath-like upon velocity-reversal: instead it meticulously unravels its correlations with X. Replacing the bath with a newly prepared one of equal temperature would erase the echo.
\\

\noindent{\bf Further implications}:

We have constructed a Microscopic Dynamical Entropy (MDE), $S^{\rm md}_{XY}(t)$, which matches quantitatively the time-dependent thermodynamic entropy $S^{\rm th}_{XY}(t)$ of a system X and its heat bath Y, with the XY system obeying Hamiltonian dynamics. The MDE is neither the Gibbs entropy $S^G_{XY}$, which is conserved, nor its coarse-graining via phase-cells $S^{\rm cg}_{XY}(t)$, which can never decrease at all. Both of these depend on dynamics within the heat bath, while the MDE explicitly does not. Our MDE not only matches by construction the entropy of classical thermodynamics but also matches the one studied in stochastic thermodynamics. This is natural, as the latter's separation into system and environment already assumes  that Y interacts with X on `bath-like' terms~\cite{Seifert2012}.

The MDE is the Gibbs entropy under a coarse-graining that nondestructively ignores the detailed dynamics of the bath. It shows a second law under normal conditions; here the X dynamics are insensitive to bath details, while the latter become ever more complicated over time. However, atypical ensembles exist for which the reverse is true, correlations unravel, and $S^{\rm md}$ decreases. Such ensembles arise with echo protocols, realizable experimentally for spin systems, and which even when unrealized offer a stretching logical test of irreversibility theories~\cite{Ridderbos1998}.

Arguably, the very concept of a heat bath, a cornerstone of classical thermodynamics, already breaks time reversal symmetry. An arrow of time emerges because, when a system X is first put in contact with a heat bath Y to form an isolated XY system, the two are not already strongly correlated, as they will later become. A consistent MDE should thus embody the heat bath concept in its coarse-graining scheme, as ours does. 

Finally, much of our formalism carries over when bath Y is not much larger than subsystem X. Conceivably, two similar parts X and Y of an isolated system could each act as a bath for the other (albeit sacrificing time-scale separation). This may point to an MDE for isolated systems that are not of XY type. However, although the second law is concisely stated as ``$dS^{\rm th}/dt\ge 0$ for an isolated system'', thermodynamics is the theory of heat. A full statistical interpretation might not be possible without appeal to the heat-bath concept.
\\

{\em Acknowledgements:} We thank Xiangjun Xing for important discussions in the early stages of this work. We thank Michael te Vrugt for a most interesting seminar on irreversibility problems.

\end{document}